\documentclass[aps,epsfig,preprint]{revtex4-1}
\usepackage{color} 
\usepackage{xcolor}
\usepackage{graphicx} 
\usepackage{epsfig}

\newcommand{\bee}{\begin{equation}}
\newcommand{\ene}{\end{equation}}
\newcommand{\beea}{\begin{eqnarray}}
\newcommand{\enea}{\end{eqnarray}}

\begin{document}
\title{Relativistic electron beam driven longitudinal wake-wave breaking in a cold plasma }
\author{Ratan Kumar Bera, Arghya Mukherjee, Sudip Sengupta and Amita Das}
\affiliation{Institute for Plasma Research, Bhat , Gandhinagar - 382428, India }
\date{\today}
%{for revtex  maketitle should be written just here}
%\maketitle 
\begin{abstract}  
 Space-time evolution of relativistic electron beam driven wake-field in a cold, homogeneous plasma, 
 is studied using 1D-fluid simulation techniques.
 It is observed that the wake wave gradually evolves and eventually breaks, exhibiting sharp spikes in the density profile and sawtooth like features in the electric
 field profile \cite{ratan}. It is shown here that the excited wakefield is a longitudinal Akhiezer-Polovin mode \cite{Akhiezer} and its steepening (breaking) can be understood in terms of phase 
 mixing of this mode, which arises because of relativistic mass variation effects. Further the phase mixing time (breaking time) is studied as a function of 
 beam density and beam velocity and is found to follow the well known scaling presented in ref.\cite{arghya}.

\end{abstract}
% for revtex4 here maketitle should be written
\maketitle 
%\begin{multicols}{2}
\section{ Introduction}

Plasma based acceleration schemes have shown promising results in recent years \cite{modena,muggli,santala,surf}. 
Plasmas form an attractive medium for future generation of accelerators, because they can support electric fields of the order of several hundred $GV/m$,
which is many orders of magnitude higher than that produced by conventional RF based accelerators \cite{malkascience}. These extreme fields are generated 
by relativistically intense longitudinal plasma waves, which are excited when an ultra-intense laser pulse or an ultra-relativistic beam pulse propagates
through the plasma \cite{esarey,Hooker,cj,Faure2006,chan,Muggli2009116,chen,uhm,katsouleas}. Based on the mechanism of excitation of the plasma wave (wake wave),  
plasma based acceleration schemes are categorized into two types, Laser Wakefield Acceleration (LWFA) and Plasma Wake-field Acceleration (PWFA).   
In Laser Wakefield Acceleration (LWFA) scheme, an ultrashort, intense laser pulse is employed to drive a relativistically intense plasma wave. 
Charged particles either externally injected or trapped from the background plasma ride on this excited plasma wake wave and get accelerated 
to high energies.
This acceleration process has been confirmed in a number of experiments by accelerating electrons to GeV energies \cite{malkalaser,Golovin,malka}.
In Plasma Wakefield Acceleration (PWFA) scheme, an intense, near light-speed electron beam is used instead of 
a laser pulse to excite plasma wave which has a phase velocity equal to the velocity of the beam.
A late coming bunch of charged particles rides on this wave and gets accelerated to
high energies. 
As ``plasma afterburners'' this scheme is most suitable to boost the energy of the existing linacs.
In 2007, Blumenfeld et al. \cite{Blumenfeld2007}  have accelerated electrons from the tail of a driver bunch having energy 42 GeV 
up-to a maximum energy of 85 GeV, in a meter long plasma at SLAC (Stanford Linear Accelerator Center). 
But the accelerated electrons had a very broad energy spectrum. Recently in 2014, Litos et al. \cite{litos} 
have also demonstrated the success of PWFA scheme achieving a much lower spread of accelerated beam energy (hardly 2 percent) by injecting 
a discrete trailing bunch. 
\par\vspace{\baselineskip}
The structure of the wakefield  excited by an ultra-relativistic electron beam pulse propagating through a plasma, has been studied extensively by 
Rosenzweig et al. \cite{rosenzweig}, Amatuni et al. \cite{amatuni} and Ruth et al. \cite{ruth}. Due to intrinsic interest in their non-linear properties 
numerous investigations have been carried out, both numerically and analytically, in this area. In an earlier study, Rosenzweig et al. \cite{rosenzweig}
gave an analytical expression, in 1-D, for the wake electric field excited by an ultra-relativistic electron beam having density ($n_b$) less than
or equal to half the plasma density ($n_0$). In our recent work \cite{ratan}, we reported a detailed analytical and numerical study of 
relativistic electron beam driven wakefield, where we analytically extended Rosenzweig's work to arbitrary beam densities and numerically verified our
analytical results using 1-D fluid simulation. Our simulation result exhibited a good match with the analytical results for several plasma periods.
However, it was observed that at late times in the simulation, the perturbed density in all cases show spiky features, which is accompanied by sawtooth
like structures in the electric field profile. This particular behavior was absent in our analytically derived 
profile of perturb density and electric field (see ref. \cite{ratan}).
The spiky features in the perturb density profile and sawtooth like structures in the electric field profile 
are well known signatures of wave-breaking \cite{infeld,sudip_pre}.
In our previous report, we had stated that the excited wake wave
is a longitudinal Akhiezer-Polovin wave \cite{Akhiezer} which
breaks when perturbed longitudinally \cite{prabal}. In our
case, perturbation is produced by numerical noise. 

\par\vspace{\baselineskip}
In this paper, we have extended our earlier work and present a detailed study of the breaking of wake wave and its dependence on the electron beam density
and velocity. We have followed the space-time evolution of the electron beam driven wake wave in a cold plasma using 1-D fluid simulation. We have carried 
out the simulation for a long enough time for the wake wave to break and exhibit spiky features. In section II, we present the equations governing the 
evolution of wake field. To clearly study the breaking of wake wave, we have neglected the beam evolution in the self consistent field of the wake wave.
This is valid in the limit $\gamma_b \gg 1$, where $\gamma_b$ is the Lorentz factor associated with the beam velocity ($v_b$). Section III contains a 
brief discussion of the simulation techniques and our simulation results. It is observed that the wake wave evolves in time and breaks after several plasma 
periods. In section IV, the physical mechanism underlying wake wave breaking is discussed and the numerical results are compared with the well known scaling 
of Akhiezer-Polovin wave breaking time with phase velocity of the wake wave ($\beta_{ph}$) and maximum fluid velocity ($u_m$) \cite{arghya}.

\par\vspace{\baselineskip}
\section{Governing Equations}
The basic equations governing the space and time evolution of ultra-relativistic electron beam driven wakefield in a cold plasma are the relativistic 
fluid-Maxwell equations for the plasma electrons. 
As stated in the introduction, we work in the limit of $\gamma_b \gg 1$, where the beam evolution equations may be neglected. Also ion dynamics is neglected,
as ions do not respond in these time scales. 
Ions are only assumed to provide a stationary neutralizing background.
We consider beam to be moving along $z$-direction in an infinite, homogeneous cold plasma. 
Neglecting the variation of plasma parameters in the transverse 
(transverse to the beam propagation) direction, the basic governing equations are the continuity and momentum equation for plasma electrons and Poisson's
equation, which in normalized form in 1-D are given as  
\begin{equation}
  \frac{\partial n}{\partial t}+\frac{\partial (nv) }{\partial z}=0	\label{pl_cont}
 \end{equation}
 \begin{equation}
  \frac{\partial p }{\partial t}+v\frac{\partial p}{\partial z}=-E      \label{pl_mom}
 \end{equation}

 \begin{equation}
\frac{\partial E }{\partial z}=(1-n-n_b) \label{pois}
 \end{equation}
 where $p=\gamma v$ is the $z$-component of momentum of plasma electrons having 
 $z$-component of velocity $v$ and $\gamma=\left(1-v^2\right)^{-1/2}$ is the relativistic 
 factor for plasma electrons. 
 $E$ and $n_0$ are the $z$-component of the self-consistent electric field and equilibrium plasma density respectively.
Here we have used the normalization factors as,
$t \rightarrow \omega_{pe}t$, $z \rightarrow \frac{\omega_{pe}z}{c}$, $E \rightarrow \frac{eE}{m_e c\omega_{pe}}$, $v\rightarrow \frac{v}{c}$, 
$p \rightarrow \frac{p}{m_e c}$, $n\rightarrow \frac{n}{n_0}$, $n_b\rightarrow \frac{n_b}{n_0}$, $\omega_{pe}$ being the non-relativistic plasma
frequency and $n_0$ is the equilibrium plasma density. The above equations (equation (\ref{pl_cont}-\ref{pois})) are the main key equations required to study 1-D electrostatic relativistic 
electron beam driven wakefield excitation in a cold plasma. 

\section{ Fluid simulation of relativistic electron beam driven wakefield}
\begin{figure}
    \includegraphics[width=0.8\textwidth]{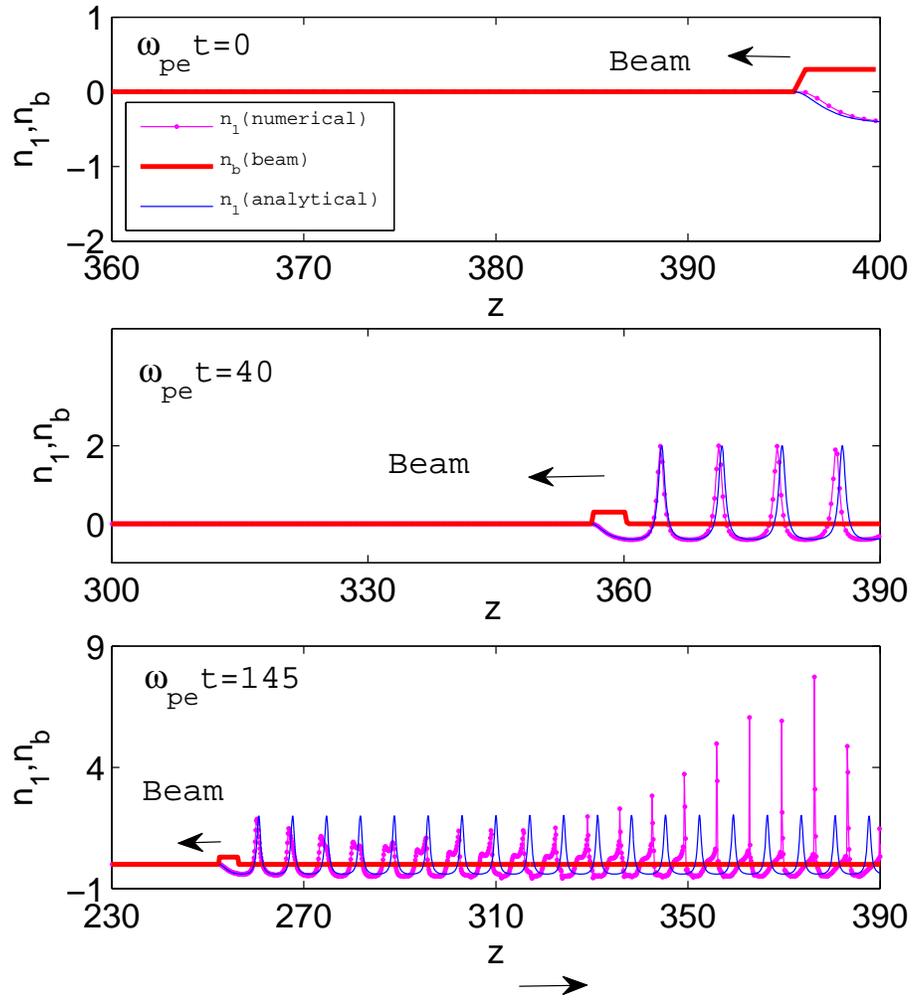}
    \caption{Numerical and analytical normalized perturbed electron density ($n_1$) profile at different times for the normalized beam density ($n_b$)=0.3 and 
    beam velocity ($v_b$) =0.99}
    \label{fig1}
\end{figure}

\begin{figure}
    \includegraphics[width=0.8\textwidth]{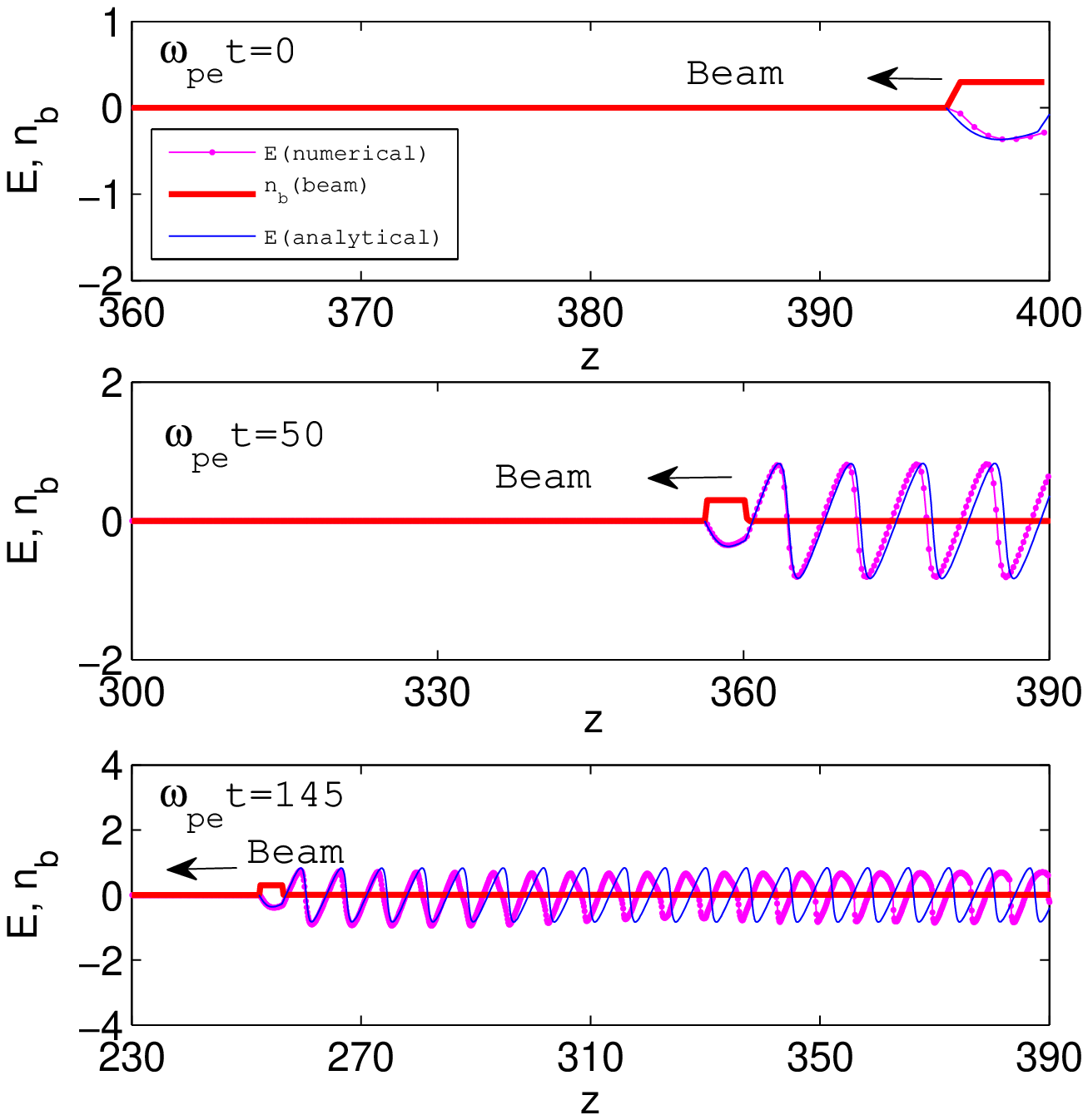}
       \caption{Numerical and analytical normalized electric field ($E$) profile at different times for the normalized beam density ($n_b$)=0.3 and 
    beam velocity ($v_b$) =0.99}
       \label{fig2}
\end{figure}

\begin{figure}
    \includegraphics[width=0.8\textwidth]{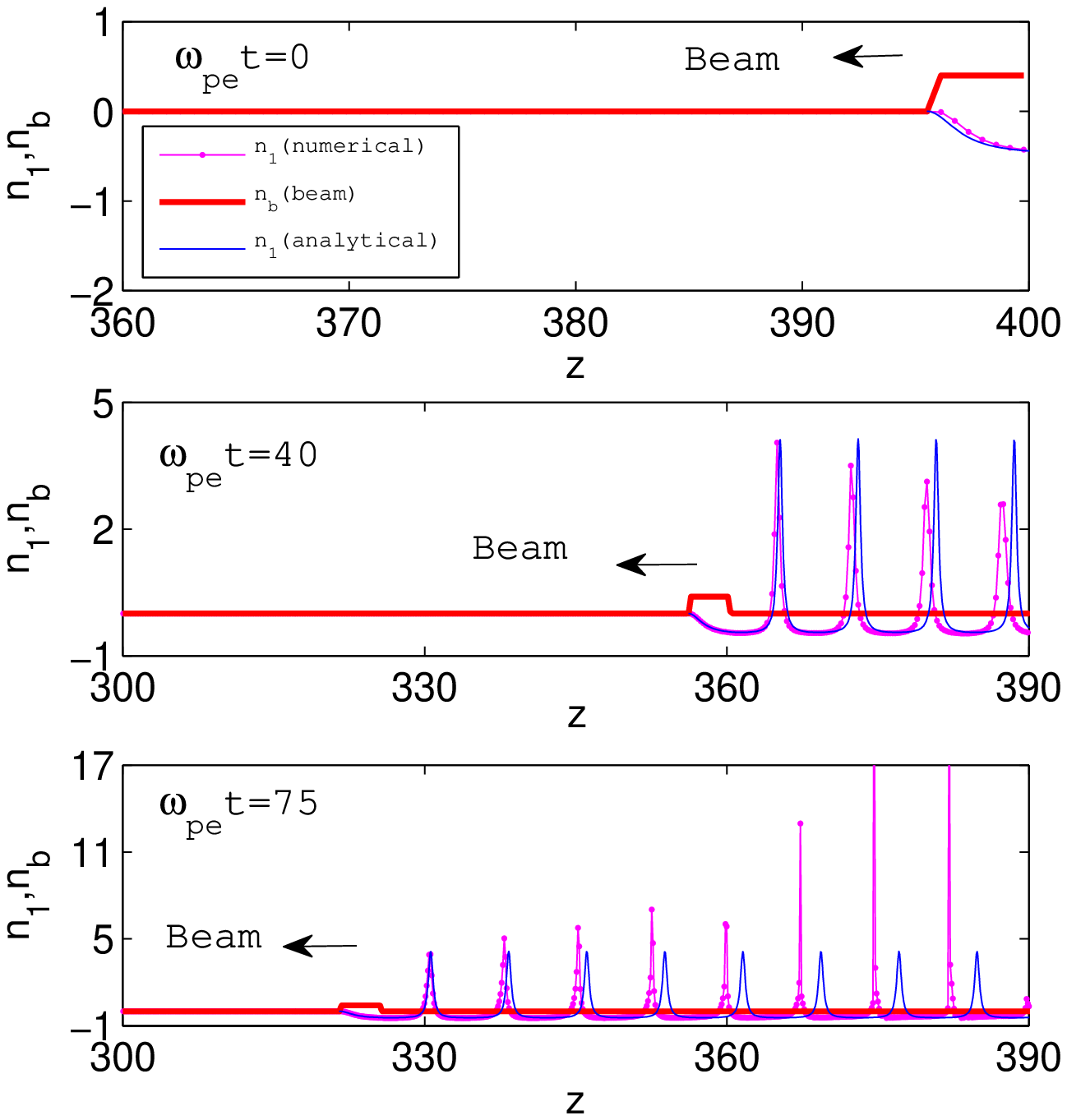}
    \caption{Numerical and analytical normalized perturbed electron density ($n_1$) profile at different times for the normalized beam density ($n_b$)=0.4 and 
    beam velocity $v_b=0.99$}
    \label{fig3}
\end{figure}

\begin{figure}
    \includegraphics[width=0.8\textwidth]{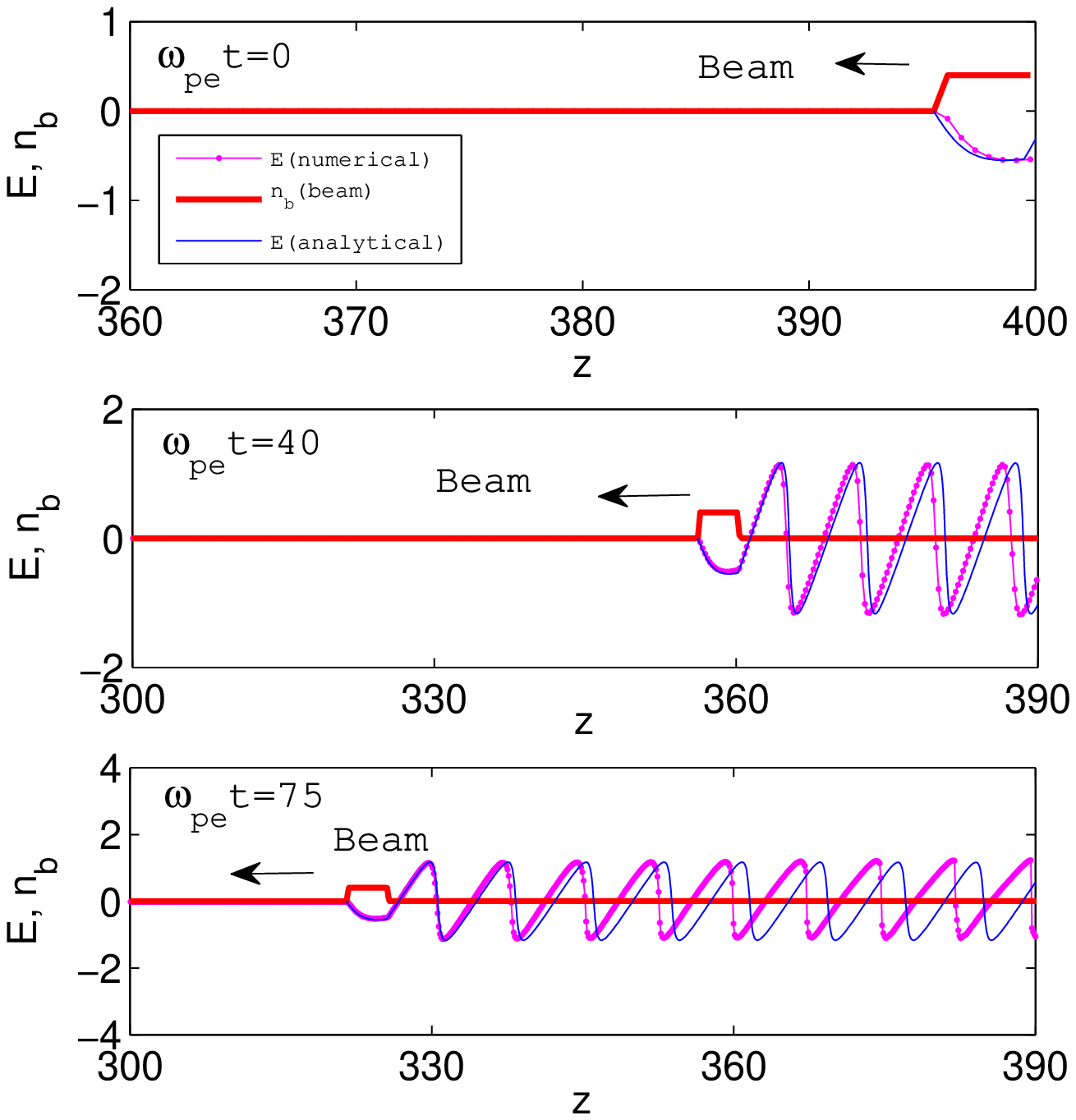}
       \caption{Numerical and analytical normalized electric field ($E$) profile at different time for the normalized beam density ($n_b$)=0.4
       and beam velocity $v_b=0.99$}
       \label{fig4}
\end{figure}

\begin{figure}
    \includegraphics[width=0.8\textwidth]{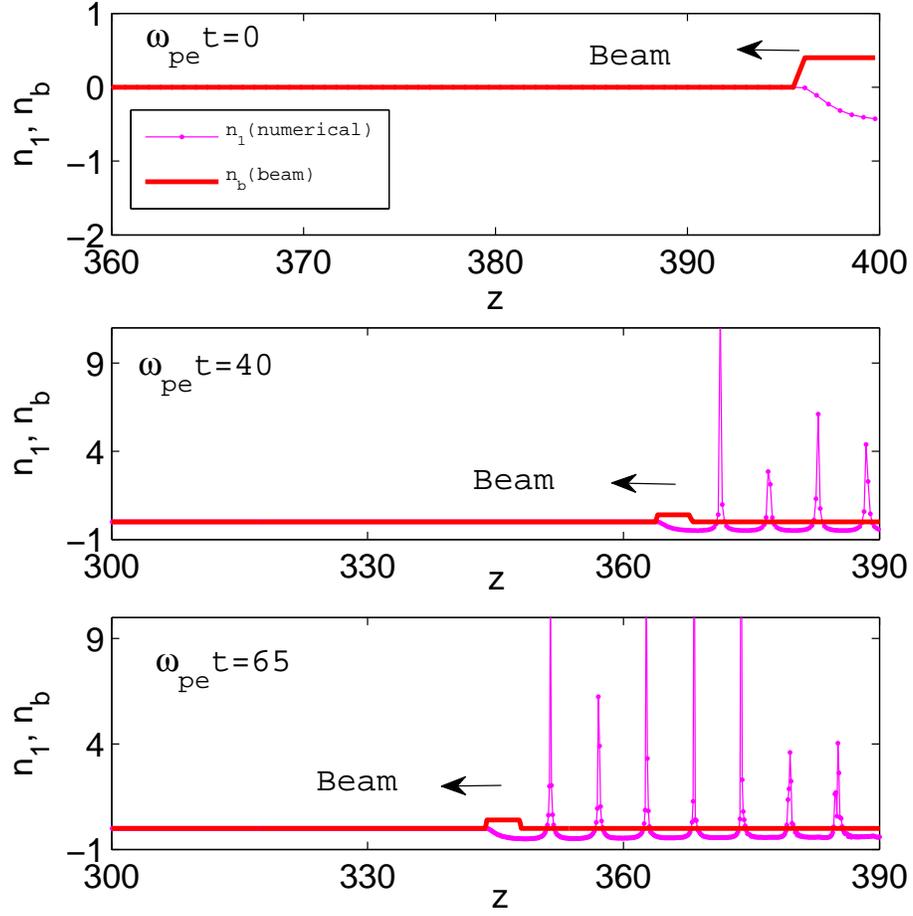}
     \caption{Numerical normalized perturbed electron density ($n_1$) profile at different times for the normalized beam density ($n_b$)=0.4 and 
    beam velocity $v_b=0.8$}
       \label{fig5}
\end{figure}

\begin{figure}
    \includegraphics[width=0.8\textwidth]{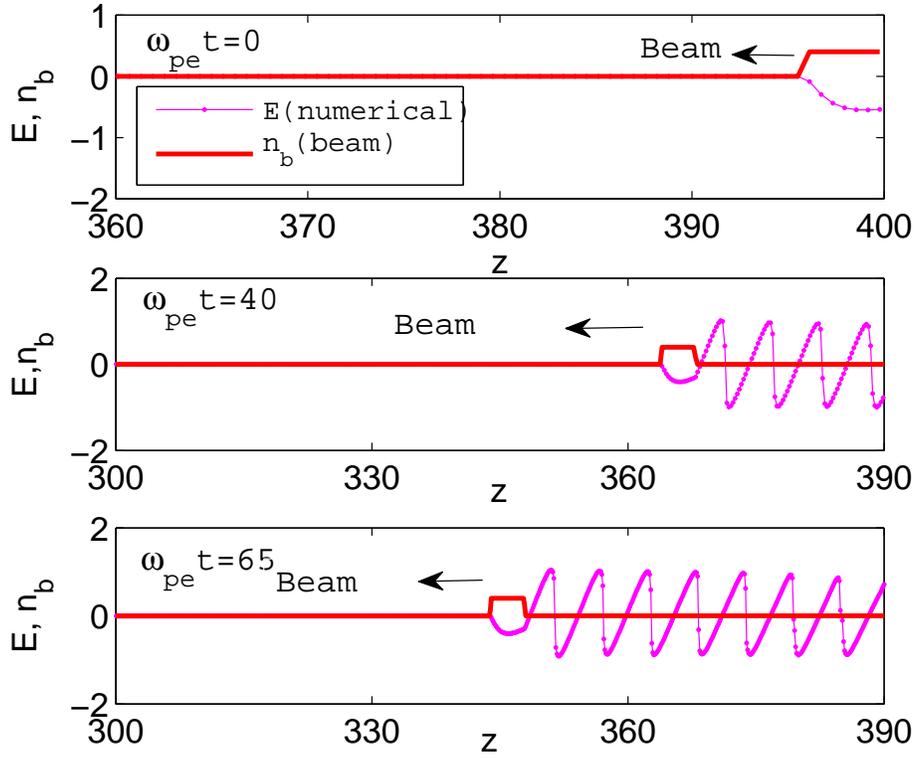}
       \caption{Numerical normalized electric field ($E$) profile at different times for the normalized beam density ($n_b$)=0.4 and 
    beam velocity $v_b=0.8$}
       \label{fig6}
\end{figure}

 %
 %
 %%%%
In this section, we briefly discuss the numerical techniques used to study the relativistic electron beam driven wakefield excitation in a cold plasma
and present our simulation results. We have developed a 1D fluid code using a set of subroutines (LCPFCT) which is based on flux-corrected transport 
scheme \cite{boris}, to study the space and time evolution of an ultra-relativistic electron beam 
driven wakefield in a cold plasma. The principle of this scheme is based on the generalization of 
two step Lax-Wendroff method \cite{numr}. 
We have simulated equations (\ref{pl_cont}),(\ref{pl_mom}) and (\ref{pois}) 
with non-periodic boundary conditions. Beam is considered to be rigid.
We have initiated the simulation using the profiles of electric field, density and velocity from the analytical work of
Rosenzweig et al. \cite{rosenzweig}, although our results are independent of the initial choice of profiles.
Here the driver beam is allowed to propagate inside the plasma starting from one end of the simulation window and the wake field is evolved according 
to equations (\ref{pl_cont}),(\ref{pl_mom}) and (\ref{pois}) \cite{ratan}.  
The simulation results are shown in figures(\ref{fig1}-\ref{fig6}) for different values of beam density ($n_b$) and beam velocity ($v_b$).
Fig. (\ref{fig1}) and (\ref{fig2}) respectively show the perturbed electron density and the wake electric field  for $n_b=0.3$ and $v_b=0.99$. 
Same quantities are shown in figures (\ref{fig3}) and (\ref{fig4}) for a different beam density $n_b=0.4$, keeping the beam velocity fixed 
(at $v_b=0.99$). Finally figures (\ref{fig5}) and (\ref{fig6}) respectively show the perturbed electron density and wake electric field for $n_b=0.4$
and a different beam velocity $v_b=0.8$. In all the figures, numerical results are shown in magenta and the analytical results (derived in ref. \cite{ratan}
and reproduced in the next section, for completeness) are shown in blue.

\par\vspace{\baselineskip}

\begin{figure}[h]
    \includegraphics[width=0.8\textwidth]{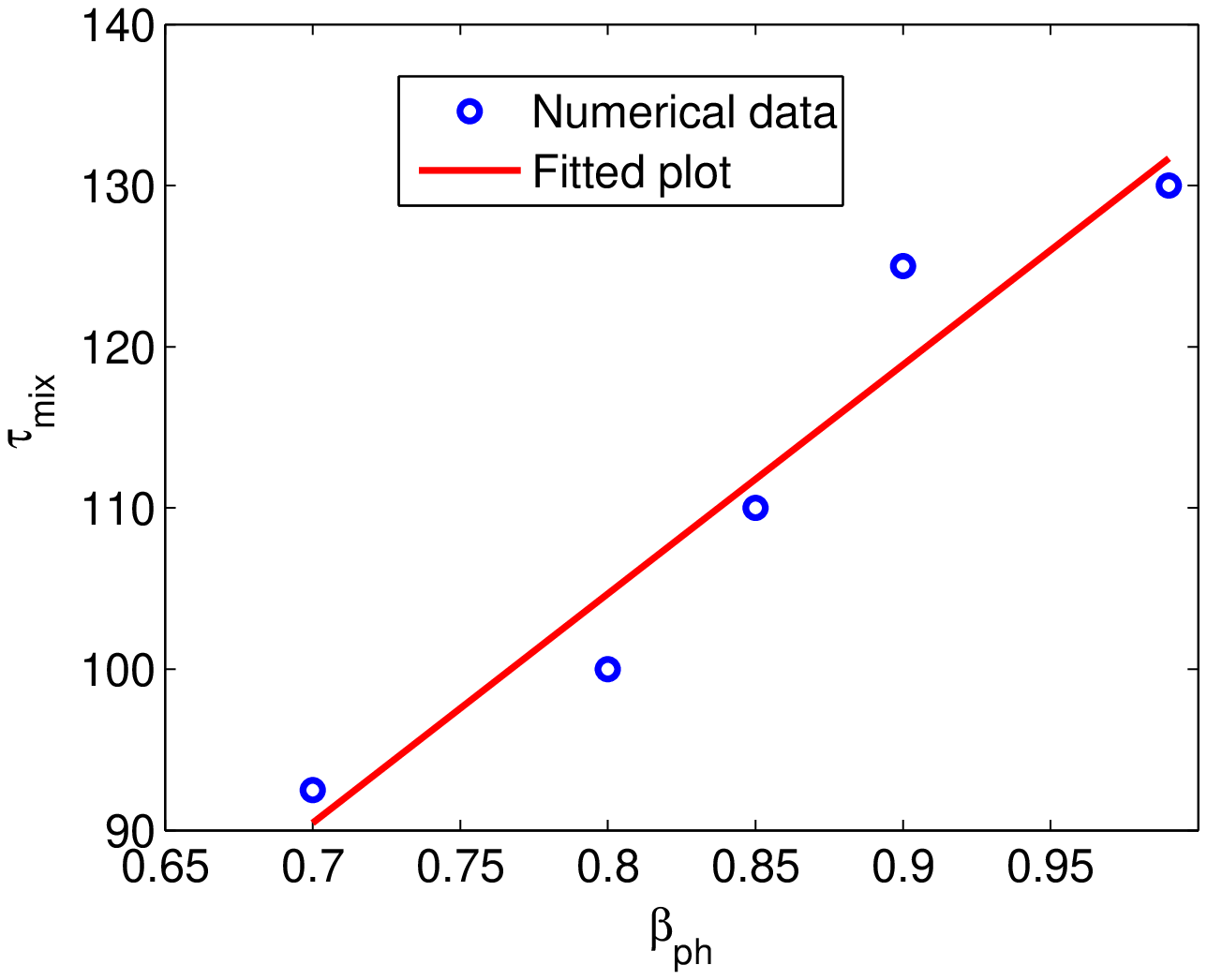}
    \caption{Plot for numerically obtained (circles) and fitted (solid) scaling of Phase mixing time ($\tau_{mix}$) with the phase velocity ($\beta_{ph}$) 
    for the normalized beam density ($n_b$)=0.3}
    \label{fig7}
\end{figure}

\begin{figure}[h]
    \includegraphics[width=0.8\textwidth]{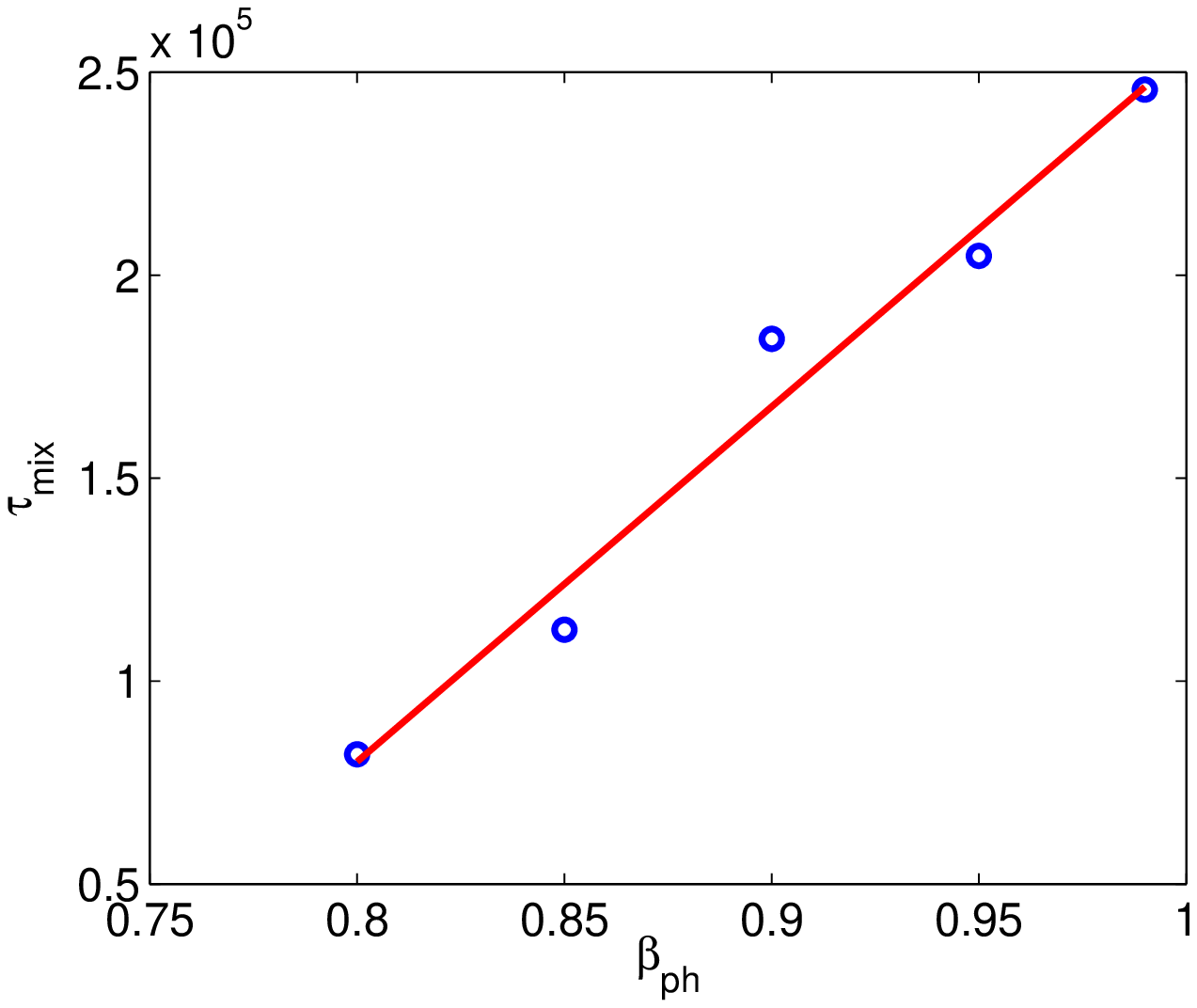}
   \caption{Plot for numerically obtained (circles) and fitted (solid) scaling of Phase mixing time ($\tau_{mix}$) with the phase velocity ($\beta_{ph}$) 
    for the normalized beam density ($n_b$)=0.4}
    \label{fig8}
\end{figure}

\begin{figure}[h]
    \includegraphics[width=0.8\textwidth]{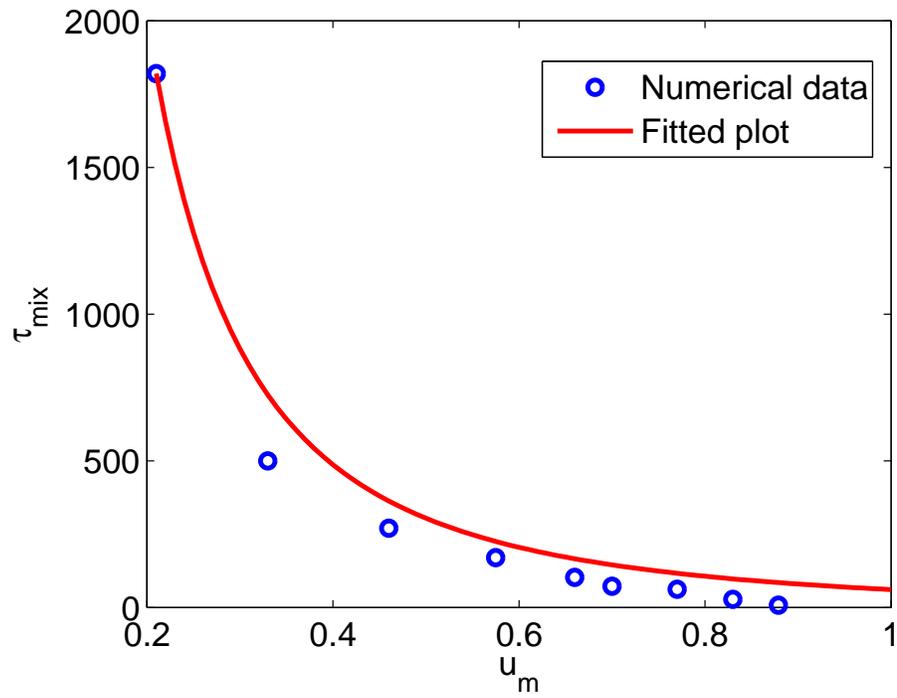}
    \caption{Plot for analytical (solid) and numerical (circles) scaling of Phase mixing time ($\tau_{mix}$) as a function of maximum fluid velocity ($u_m$).}
    \label{fig9}  
\end{figure}

As  mentioned in the introduction and also in ref. \cite{ratan}, it is observed that in all cases the simulated wakefield profile gradually 
deviates from analytical profile, with time and eventually breaks after several plasma periods.
The signature of breaking of the wake wave is seen as density spikes in the wake wave. The electric field also 
exhibits into a sawtooth structure close to breaking time.
It is observed that for a fixed beam velocity, higher the beam density shorter is the wake wave breaking time; whereas for a fixed beam density,
higher the beam velocity longer is the wake wave breaking time. Fig. (\ref{fig7}) and (\ref{fig8}) show the variation of breaking time with 
the beam velocity (phase velocity of the wake wave) for fixed beam densities $n_b=0.3 $ and 0.4 respectively and fig. (\ref{fig9}) shows the variation 
of wake wave breaking time with the max. fluid velocity ($u_m$) (``$u_m$'' is related to the beam density, as discussed in the next section).
In these figures (\ref{fig7}, \ref{fig8} and \ref{fig9}), the points are obtained from simulation and continuous lines present our understanding of 
the breaking mechanism of wake wave in terms of breaking of longitudinal Akhiezer-Polovin wave. This is presented in the next section.

%
%========================================================================
\section{Analysis of wake wave breaking}

In this section, we present a detailed discussion on the physical mechanism of wake wave breaking and the dependence of wake wave 
breaking time on the electron beam velocity and density. For the sake of completeness we first present the analytical expression for 
wakefield profile behind the beam. In terms of wave frame variable $\tau= (t-\frac{z}{\beta_{ph}})$ and in the limit $\beta_{ph} \rightarrow 1$,
the wake wave profile behind the beam may be written in parametric form as

\begin{equation}
 E=\left(\frac{\gamma_m^2 -1}{\gamma_m +\sqrt{\gamma_m^2 -1 } cos(2\phi)}\right)^{\frac{1}{2} }sin (2\phi) \label{wakeelec}
\end{equation}

\begin{equation}
\tau=\tau_f+ 2\sqrt{\gamma_m +\sqrt{\gamma_m^2 -1}}[E(\phi_f,m)-E(\phi,m)] \label{tauw}
\end{equation}

where $\tau_f=\frac{\omega_{pe} l_b}{c}$, $l_b$ being the length of the beam and $\gamma_m$ and $\phi_f$ are constants; 
$\gamma_m =(1-u_m^2)^{-\frac{1}{2}}$ is the Lorentz factor associated with the maximum fluid velocity ``$u_m$'' behind 
the beam and $\phi_f$ is the value of the parameter $\phi$ at $\tau=\tau_f$. $E(\phi,m)$, $E(\phi_f,m)$ are incomplete elliptic integral of second kind with
$m^2=2(\sqrt{\gamma_m^2-1})/(\gamma_m+\sqrt{\gamma_m^2-1})$. The perturbed density is given by $n_1=\frac{1}{2}\frac{1-x^2}{x^2}$ where $x$ is defined as
$x=(\frac{1-\beta}{1+\beta})^{1/2}$, $\beta$ being the fluid velocity; $x$ is related to the parameter $\phi$ as $x=\gamma_m+(\sqrt{\gamma_m^2-1}) cos(2\phi)$.
Thus all the wake field variables, perturbed density, electric field, and fluid velocity are represented in terms of the parameter $\phi$. The constants
$\gamma_m$ and $x_f$ (value of $x$ corresponding to $\phi_f$) are evaluated using the beam density $\alpha(=n_b)$ and length $l_b$ (or $\tau_f$) as

\begin{equation}
 \gamma_m=(1-\alpha) +\alpha x_f \label{gamma}
\end{equation}

where $x_f$ is related to $\tau_f$ through the implicit relations
\begin{equation}
 x_f=\frac{1-2\alpha sin^2 \psi_f}{1-2\alpha}  \label{xf}
\end{equation}
and
\begin{equation}
\tau_f=2(1-2\alpha)^{-1}[E(k)-E(\psi_f,k_1)] \label{tau1}
\end{equation}

Here $E(\psi_f,k_1)$ and $E(k_1)$ are respectively the incomplete and complete elliptic integrals of second kind with $k_1^2 =2\alpha$. Equations (\ref{gamma}), 
(\ref{xf}) and (\ref{tau1}) are derived using wakefield equations inside the beam and using the continuity conditions at the end of the beam 
(for complete details see ref. \cite{ratan}).  Equation (\ref{xf}) and (\ref{tau1}) are valid for $\alpha< 1/2$, which is the range of beam density 
within which we have limited our present set of simulation. The frequency of the wake wave behind the beam is given by 
$\omega_{wake}=\pi \left(2\sqrt{(\gamma_m+\sqrt{\gamma_m^2-1})}E(m)\right)^{-1}$. Using equations (\ref{gamma}-\ref{tau1}) for a given beam
density $\alpha$ and beam length $l_b$, the wakefield profiles behind the beam (perturbed electron density and electric field) are plotted along with the 
simulation results in Fig. (\ref{fig1})-(\ref{fig4}). In figure (\ref{fig5}) and (\ref{fig6}), the simulation results are compared with the numerical solution 
of the wakefield differential equations, $\beta_{ph}\neq 1$ (see ref. \cite{ratan}). 
As mentioned earlier, simulation results match well with the analytical expressions for several plasma periods, 
but at late times a marked deviation between the two is observed. Sharp spikes in perturbed density is accompanied with 
sawtooth profile in wake electric field. These features are well known signature of wake wave breaking.
In order to understand this phenomenon, we first identify the wake wave with a longitudinal Akhiezer-Polovin mode.

\par\vspace{\baselineskip}

\begin{figure}[h]
    \includegraphics[width=0.8\textwidth]{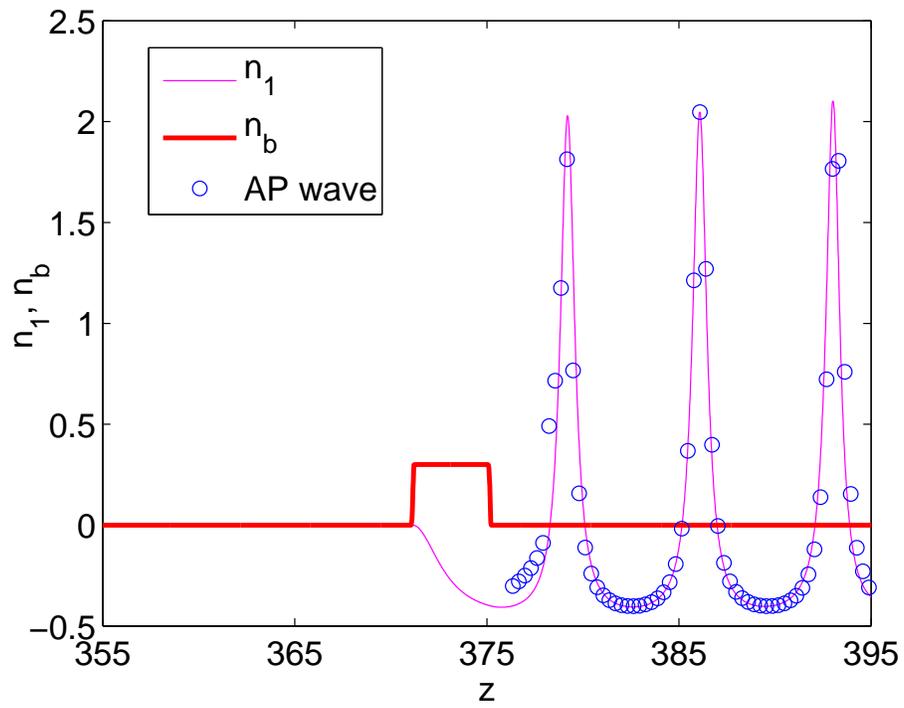}
    \caption{Plot of perturbed density profile ($n_1$) of wake wave (magenta) and Akhiezer-plovin mode (blue circles) }
    \label{fig10}  
\end{figure}

It is well known that the stationary wave frame solution of the relativistic fluid-Maxwell equations in 1-D,
for a cold homogeneous plasma with infinitely massive ions (equations (\ref{pl_cont})-(\ref{pois}) without the beam terms in the Poisson's equation) is
a longitudinal Akhiezer-Polovin mode which is parameterized in terms of maximum fluid velocity ``$u_m$''. and phase velocity ``$\beta_{ph}$'' 
\cite{Akhiezer,prabalpop}.  Thus the wakefield behind the beam, which is a solution of equations (\ref{pl_cont})-(\ref{pois}) with $\alpha =0$, 
is nothing but a longitudinal Akhiezer-Polovin mode, where the parameter $u_m$ is related to the beam density $\alpha$ and the length of the beam $l_b$
through the equations (\ref{xf}-\ref{tau1}). Also using the identity $E(\frac{2\sqrt{k}}{1+k})= \frac{1}{1+k}\left(2E(k)-k'^2 K(k)\right)$ \cite{ryzhik}, with
$k^2=\frac{\gamma_m -1}{\gamma_m +1}$ and $k'^2 =1-k^2$, the expression for wake becomes identical with the 
expression for Akhiezer-Polovin wave frequency (equation (11) of ref. \cite{prabal} and equation (5) of ref. \cite{arghya} ). To further emphasize the equivalence 
between the wake wave excited by an ultra-relativistic electron beam with beam density $\alpha=n_b$ and length $\tau_f$ (or $l_b$), and a 
longitudinal Akhiezer-Polovin mode with parameter ``$u_m$'' and ``$\beta_{ph}$'', we first estimate ``$u_m$'' for the parameters of fig. (\ref{fig1}) 
using equations (\ref{gamma}-\ref{tau1}). Using this value of $u_m$ and $\beta_{ph}=\beta_b \rightarrow 1$, and following the method outlined in refs. 
\cite{prabal,arghya}, we plot the appropriate Akhiezer-Polovin mode along with the wakefield behind the beam. This is shown in fig. (\ref{fig10}), 
which clearly establishes that the two are identical.

\par\vspace{\baselineskip}

It is well known that the amplitude of a Akhiezer-Polovin mode is limited by the wave breaking limit which is given by $E_{WB}= \sqrt{2(\gamma_{ph}-1)}$,
where $\gamma_{ph}=(1-\beta_{ph}^2)^{-1}$ is the Lorentz factor associated with the phase velocity of the wave.  
For $\beta_{ph} \rightarrow 1$, wave breaking limit $E_{WB} \rightarrow \infty$, and the mode in-principle should never break. But this is contrary to what 
is observed in our simulations; the wake wave breaks at a much lower amplitude. Recently it has been shown that an Akhiezer-Polovin mode can break at
an amplitude well below its wave breaking limit via a process called phase mixing, when it is subjected to an arbitrary small longitudinal perturbation
\cite{prabal}. It has been shown in ref. \cite{arghya} that addition of an arbitrary 
small longitudinal perturbation to a longitudinal Akhiezer-Polovin mode results in the frequency of the mode acquiring a spatial dependence 
due to relativistic mass variation effects. In the present case, perturbation arises due to numerical noise. Because of the spatial dependence in frequency,
different ``pieces'' of the wave slowly go out phase with each other as time progresses. The process of phase mixing is clearly visible in the electric
field profile, where the phase difference between simulated wake field and analytically obtained wake field slowly increases with time. Phase mixing 
eventually leads to breaking of the wake wave. 
The phenomenon of phase mixing leading to wave breaking of a relativistically intense longitudinal wave have been studied extensively
by several author in different contexts  \cite{sudip_pre,infeld,sudip_ppcf,chandan}.
It is shown in ref. \cite{arghya}, that the time in which wake breaks scales with the phase velocity
$\beta_{ph}$ and maximum fluid velocity $u_m$ as $\tau_{mix} \sim \frac{2\pi}{3}\frac{\beta_{ph}}{\delta}\left(\frac{1}{u_m^2} -\frac{1}{4}\right)$, where 
``$\delta$'' is the amplitude of the perturbation. We have verified this scaling in our simulations by first keeping ``$u_m$'' fixed (i.e. $\alpha=n_b$ 
and $l_b$ fixed) and varying $\beta_b =\beta_{ph}$, and then keeping $\beta_b=\beta_{ph}$ fixed and varying $u_m$ (i.e. by varying $\alpha$). Note
the continuous lines in fig. (\ref{fig7}), (\ref{fig8}) and (\ref{fig9}); the scaling of phase mixing time $\tau_{mix}$ with $\beta_{ph}$ and $u_m^2$
compares well with out simulation results.

%%
%%%%%%%
\section{ summary}
We have studied space-time evolution of relativistic electron beam driven wake wave in a cold homogeneous plasma using 1-D fluid simulation. 
It is found that at times, which depend of on the electron beam density and velocity, the wake wave breaks via a phenomenon called phase mixing. The 
wake wave is further identified with a longitudinal Akhiezer-Polovin mode and its breaking time scales with phase velocity ($\beta_{ph}$) and maximum
fluid velocity ($u_m$) according to a relation as suggested in ref. \cite{arghya}.

%
%%%%%%%%%================================================================================================================
%\bibliographystyle{unsrt}
%\bibliography{ratan_wake}

\begin{thebibliography}{10}
%%%%%%%%%%%%%%%%%%%%%%%

\bibitem{ratan}
Ratan Kumar Bera, Sudip Sengupta and Amita Das.
\newblock Fluid simulation of relativistic electron beam driven wakefield in a cold plasma 
\newblock {\em Phys. of Plasmas}, 22, 073109 (2015).




\bibitem{Akhiezer}
A.~I. Akhiezer and R.~V. Polovin.
\newblock Theory of Wave Motion of an Electron Plasma.
\newblock {\em Sov. Phys. JETP}, 3, 696(1956).


\bibitem{arghya}
Arghya Mukherjee and sudip Sengupta.
\newblock Analytical estimate of phase mixing time of longitudinal Akhiezer-Polovin waves.
\newblock {\em Phys. of Plasmas}, 21, 112104 (2014).


\bibitem{modena}
A. Modena, Z.Najmudin, A. E.Dangor, C. E.~Clayton, K. A.Marsh, C.Joshi, V.Malka, C. B.Darrow, C. Danson, D. Neely, F. N.~Walsh.
\newblock Electron acceleration from the breaking of relativistic plasma waves.
\newblock {\em Nature }, 377, 606-608 (1995).

\bibitem{muggli}
P. Muggli, B. E. Blue, C. E. Clayton, S. Deng, F.-J. Decker, M. J. Hogan, C. Huang, R. Iverson, C. Joshi, T. C. Katsouleas, S. Lee,
W. Lu, K. A. Marsh, W. B. Mori, C. L. O'Connell, P. Raimondi, R. Siemann, and D. Walz.
\newblock Meter-Scale Plasma-Wakefield Accelerator Driven by a Matched Electron Beam.
\newblock {\em Phys. Rev. Lett. }, 93, 014802 (2004).

\bibitem{santala}
M. I. K. Santala, Z. Najmudin, E. L. Clark, M. Tatarakis, K. Krushelnick, A. E. Dangor, V. Malka, J. Faure, R. Allott, and R. J. Clarke.
\newblock Observation of a Hot High-Current Electron Beam from a Self-Modulated Laser Wakefield Accelerator.
\newblock {\em Phys. Rev. Lett. }, 86, 1227 (2001).


\bibitem{surf}
Mike Downer and Rafal Zgadzaj.
\newblock Accelerator physics: Surf's up at SLAC.
\newblock {\em Nature}, 515, 40(2014).



\bibitem{malkascience}

 V. Malka, S. Fritzler, E. Lefebvre, M. M. Aleonard, F. Burgy, J.~P. Chambaret, J. F. Chemin, K. Krushelnick,
 G. Malka, S. P. D. Mangles, Z. Najmudin, M. Pittman, J.P. Rousseau, J.N. Scheurer, B. Walton, A. E. Dangor
\newblock Electron Acceleration by a Wake Field Forced by an Intense Ultrashort Laser Pulse.
\newblock {\em Science }, 298, 1596 (2002).



\bibitem{esarey}
E.~Esarey, C.~B. Schroeder, and W.~P. Leemans.
\newblock Physics of laser-driven plasma-based electron accelerators.
\newblock {\em Rev. Mod. Phys.}, 81, 1229(2009).

\bibitem{Hooker}
S.~M. Hooker.
\newblock Developments in laser-driven plasma accelerators.
\newblock {\em Nat. Photon}, 7, 775(2013).

\bibitem{cj}
C.~Joshi.
\newblock The development of laser- and beam-driven plasma accelerators as an experimental field.
\newblock {\em Physics of Plasmas}, 14, 055501(2007).

%\bibitem{benno}
%Benno Zeitler, Irene Dornmair, Tim Gehrke, Mikheil Titberidze, R.~A. Maier, Bernhard Hidding, Klaus Flöttmann and  Florian Grüner.
%\newblock Merging conventional and laser wakefield accelerators.
%\newblock {\em Proc. SPIE}, 8779, 877904(2013).



%\bibitem{laser}
%W.~Lu, M.~Tzoufras, C.~Joshi, F.~S. Tsung, W.~B. Mori, J.~Vieira, R.~A. Fonseca and L.~O. Silva.
%\newblock Generating multi-GeV electron bunches using single stage laser wakefield acceleration in a 3D nonlinear regime.
%\newblock {\em Phys. Rev. ST Accel. Beams}, 10, 061301(2007).

\bibitem{Faure2006}
J.~Faure, C.~Rechatin, A.~Norlin, A.~Lifschitz, Y.~Glinec and V.~Malka.
\newblock Controlled injection and acceleration of electrons in plasma wakefields by colliding laser pulses.
\newblock {\em Nature}, 444, 737(2006).


\bibitem{chan}
Chandrashekhar Joshi.
\newblock Plasma wake field accelerator.
\newblock {\em Scientific American}, 294, 40(2006).


\bibitem{Muggli2009116}
Patric Muggli and Mark J. Hogan.
\newblock Review of high-energy plasma wakefield experiments.
\newblock {\em Comptes Rendus Physique },10, 116(2009).


\bibitem{chen}
Pisin Chen, J.~M. Dawson, W.~Robert Huff and T.~Katsouleas.
\newblock Acceleration of electrons by the interaction of a bunched electron beam with a plasma.
\newblock {\em Physical Review letters}, 54, 693(1985).



% \bibitem{lu}
% W.~Lu, C.~Huang, M.~Zhou, M.~Tzoufras, F.~S. Tsung, W.~B. Mori,T.~Katsouleas.
% \newblock A nonlinear theory for multidimensional relativistic plasma wave wakefields.
% \newblock {\em Physics of Plasmas}, 13, 056709(2006).


\bibitem{uhm}
Han Sup uhm and Glenn Joyce.
\newblock Theory of wake‐field effects of a relativistic electron beam propagating in a plasma.
\newblock {\em Physics of Fluids B: Plasma Physics}, 3, 1587(1991).



\bibitem{katsouleas}
T.~Katsouleas.
\newblock Physical mechanisms in the plasma wake-field accelerator.
\newblock {\em Phys. Rev. A}, 33, 2056(1986).




%\bibitem{tajima}
%T.~Tajima and J.~M. Dawson.
%\newblock Experimental Observation of Plasma Wake-Field Acceleration.
%\newblock {\em Phys. Rev. Lett.}, 43, 267(1979).

\bibitem{malkalaser}
V.~Malka.
\newblock Laser plasma accelerators.
\newblock {\em Physics of Plasmas}, 19, 055501(2012).

\bibitem{Golovin}
G.~Golovin, S.~Chen, N.~Powers, C.~Liu, S.~Banerjee, J.~Zhang, M.~Zeng, Z.~Sheng and and D.~Umstadter.
\newblock Tunable monoenergetic electron beams from independently controllable laser-wakefield acceleration and injection.
\newblock {\em Phys. Rev. ST Accel. Beams}, 18, 011301(2015).


%\bibitem{Balk}
%V.~A. Balakirev, I,~V. Karas, V.~I. Karas, V.~D. Levchenko and M.~Bornatici.
%\newblock Charged particle acceleration by an intense wake-field excited in plasmas by either laser pulse or relativistic electron bunch
%\newblock {\em Laser and Particle Beams}, 22, 383(2004).

%\bibitem{fauremal}
%V.~Malka, J.~Faure, J.~R. Marquès, F.~Amiranoff, J.~P. Rousseau, S.~Ranc, J.~P. Chambaret, Z.~Najmudin, B.~Walton, P.~Mora and A.~Solodov.
%\newblock Tunable monoenergetic electron beams from independently controllable laser-wakefield acceleration and injection.
%\newblock {\em Physics of Plasmas}, 8, 2605(2001)

\bibitem{malka}
Chan Joshi and Victor Malka.
\newblock Focus on Laser- and Beam-Driven Plasma Accelerators.
\newblock {\em New J. Phys.}, 12, 045003(2010).

\bibitem{Blumenfeld2007}
Ian Blumenfeld, Christopher E. Clayton, Franz-Josef Decker, Mark J. Hogan, Chengkun Huang,
Rasmus Ischebeck, Richard Iverson, Chandrashekhar Joshi, Thomas Katsouleas, Neil Kirby, Wei Lu, Kenneth A. Marsh, Warren B. Mori, Patric Muggli, Erdem Oz, Robert H. Siemann, Dieter Walz and Miaomiao Zhou.
\newblock Energy doubling of 42 GeV electrons in a metre-scale plasma wakefield accelerator.
\newblock {\em Nature}, 445, 741(2007).

%\bibitem{ANL1}
%N.~Barov, M.~E.Conde, W.~Gai and J.~B. Rosenzweig.
%\newblock Propagation of Short Electron Pulses in a Plasma Channel.
%\newblock {\em Physical Review Lett.}, 80, 1(1998).

%\bibitem{ANL2}
%j.~B. Rosenzweig, P.~Schoessow, B.~Cole, W.~Gai, R.~Konecny, J.~Norem and J.~Simpson.
%\newblock Experimental measurements of nonlinear plasma wake fields 
%\newblock {\em Physical Review A.},39, 3(1989).




\bibitem{litos}
 M.~Litos, E.~Adli,	W.~An, C.~I. Clarke, C.~E.Clayton, S.~Corde,	J.~P. Delahaye,	R.~J. England, A.~S. Fisher, J.~Frederico, S.~Gessner,	
S.~Z. Green, M.~J. Hogan, C.~Joshi,	
W.~Lu,	K.~A. Marsh, W.~B. Mori,	
P.~Muggli, N.~Vafaei-Najafabadi,	
D.~Walz, G.~White, Z.~Wu, V.~Yakimenko and G.~Yocky.
\newblock High-efficiency acceleration of an electron beam in a plasma wakefield accelerator.
\newblock {\em Nature}, 515, 92(2014).




\bibitem{rosenzweig}
J.~B. Rosenzweig.
\newblock Nonlinear Plasma Dynamics in the Plasma Wake-Field Accelerator.
\newblock {\em Physical Review Letters}, 58, 555(1987).
%%%



\bibitem{amatuni}
A.~Ts. Amatuni, S.~Elbakram and E.~V. Sekhpessian.
\newblock {\em Yerevan Physics Institute Report No. ERFI 85-832}, 1985.


\bibitem{ruth}
R.~D.Ruth, A.~Chao, P.~L. Morton and P.~B.Wilson.
\newblock {\em particle Accelerators}, 17, 171(1985).



\bibitem{sudip_pre}
Sudip Sengupta, Vikrant Saxena, Predhiman K. Kaw, Abhijit Sen, and Amita Das,
\newblock Phase mixing of relativistically intense waves in a cold homogeneous plasma.
\newblock {\em Phys. Rev. E }, 79, 026404(2009).

\bibitem{infeld}
E. ~Infeld and G. ~Rowlands,
\newblock Relativistic bursts.
\newblock {\em Phys. Rev. Lett.}, 62, 1122 (1989)




\bibitem{prabal}
Prabal Singh Verma, Sudip Sengupta, and Predhiman Kaw.
\newblock Breaking of longitudinal Akhiezer-Polovin waves. 
\newblock {\em Phys. Rev. Lett.}, 108, 125005(2012).
%%%%%%%%%%




% \bibitem{sudip}
% Sudip Sengupta, Vikrant Saxena, Predhiman K.~Kaw, Abhijit Sen, and Amita Das.
% \newblock Phase mixing of relativistically intense waves in a cold homogeneous plasma.
% \newblock {\em Physical review E}, 79, 026404 (2009).




\bibitem{boris}
Jay.~P. Boris, Alexandra M. Landsberg, Elaine S. Oran and John H. Gardner.
\newblock LCPFCT- Flux-corrected Transport Algorithm for Solving Generalized Continuity Equations.
\newblock {\em Naval Research laboratory, Washington}, NRL/MR/6410-93-7192, 1993.


\bibitem{numr}
W.~Press, R.~Assmann, A.~Teukolsky, W.~Vetterling and Brian P. Flannery.
\newblock Numerical Recipes: The Art of Scientific Computing.
\newblock {\em Cambridge University Press}, 1992.




\bibitem{prabalpop}

Prabal Singh Verma, Sudip Sengupta and Predhiman Kaw .
\newblock Bernstein-Greene-Kruskal waves in relativistic cold plasma .
\newblock {\em Phys. Plasmas }, 19, 032110 (2012).


\bibitem{ryzhik}
I.~S.~Gradshteyn and I.~M.~ Ryzhik.
\newblock Table of Integrals, Series, and Products.
\newblock {\em Academic Press, INC.},  ISBN: 0-12-294760-6 (1980)





\bibitem{sudip_ppcf}
Sudip Sengupta, Predhiman Kaw, Vikrant Saxena, Abhijit Sen and
Amita Das.
\newblock Phase mixing/wave breaking studies of large
amplitude oscillations in a cold homogeneous
unmagnetized plasma.
\newblock {\em Plasma Phys.
Control. Fusion}, 53, 074014 (2011).


\bibitem{chandan}
Chandan Maity, Nikhil Chakrabarti and Sudip Sengupta .
\newblock Relativistic effects on nonlinear lower hybrid oscillations in cold plasma.
\newblock {\em J. Math. Phys.},  52, 043101 (2011).


\end{thebibliography}
%%%%%%%%%%%%%%%%%%%%%%%
\bibliographystyle{unsrt}
%%%%%%%%%%%%%%%%%%%

%
%----------------------------------------------------------------------

\newpage
%=========================================================================
% \begin{figure}
%     \includegraphics[width=0.5\textwidth]{wfe2.pdf}
%     \caption{Numerical normalized perturbed electron density ($n_1$) profile at different time for the normalized beam density ($n_b$)=0.2}
% \end{figure}
%%
%%
%%%
%%%%%%%
%%%%%%%%%

\end{document}